\documentclass[10pt]{article}
\usepackage{amsmath,amssymb,amsfonts}
\usepackage[english]{babel}
\makeatletter
\@addtoreset{equation}{section}
\makeatother

\def\nn{\nonumber} \def\bd{\begin{document}} \def\ed{\end{document}}
\def\ds{\documentstyle}
\let\bm=\bibitem
\newcommand{\be}{\begin{equation}}
\newcommand{\ee}{\end{equation}}
\newcommand{\bea}{\setlength\arraycolsep{2pt} \begin{eqnarray}}
\newcommand{\eea}{\end{eqnarray}}
\newcommand{\hoch}[1]{$\, ^{#1}$}
\def\p{\partial}

\title{\large {\bf Mass and angular momentum of charged rotating G\"{o}del black holes
in five-dimensional minimal supergravity}}
\date{}

\author{Jun-Jin Peng $^{1,2}$ \footnote{pengjjph@163.com}  \\ \\
\small $^1$ \sl Guizhou Provincial Key Laboratory of Radio Astronomy and
Data Processing, \\
\small \sl Guizhou Normal University,\\
\small Guiyang, Guizhou 550001, People's Republic of China\\
\small $^2$ \sl School of Physics and Electronic Science, Guizhou Normal University,\\
\small Guiyang, Guizhou 550001, People's Republic of China
}

\begin{document}

\maketitle
\vspace{20pt}

\begin{center}
\textbf{Abstract}
\end{center}
In this paper, for the sake of providing a concrete comparison between the usual
Abbott-Deser-Tekin (ADT) formalism and its off-shell extension, as well as comparing
the latter with the Barnich-Brandt-Compere (BBC) approach, we carry out these methods
to compute the mass and angular momentum of the rotating charged G\"{o}del black holes
in five-dimensional minimal supergravity. We first
present the off-shell ADT potential of the supergravity theories in arbitrary
odd dimensions, which is consistent with the superpotential via the BBC approach.
Then the off-shell generalized ADT method is applied to evaluate the
mass and angular momentum of the G\"{o}del-type black holes by including the contribution
from the gauge field. Finally, we strictly obey the rules of the original ADT formalism to
incorporate the contribution from the gauge field within the potential. With the help of the
modified potential, we try to seek for appropriate reference backgrounds to produce the mass
and angular momentum. It is observed that the ADT formalism
has to incorporate the contribution from the matter fields to yield physical charges
for the G\"{o}del-type black holes.

\textbf{Key words}: Conserved charge, G\"{o}del-type black hole, Abbott-Deser-Tekin formalism

\newpage

\section{Introduction}\label{one}

Since G\"{o}del metric in the four-dimensional Einstein gravity was found \cite{K4DGodel}, there has
been considerable interest in its analogues of various extended gravity theories
\cite{GGHPRGod,CARHe,BDGO,GodelGimH,BKGodSol,GodelBHWuGi,BaBCG3d,HarTaka,CARHNPB,LiFWL,DaMoP}.
The main reason for this is that such metrics, characterized by a so-called
G\"{o}del parameter, display several peculiar features such as
the allowance of closed timelike curves. Particularly,
within the framework of the five-dimensional minimal supergravity endowed with the
Einstein-Maxwell-Chern-Simons Lagangian, the exact G\"{o}del-type solutions that describe
the general non-extremal (charged) rotating black holes were constructed in
\cite{GodelGimH,GodelBHWuGi}.

As usual, in order to have a good understanding of the thermodynamics properties for
G\"{o}del-type solutions, it is of great necessity to identify their conserved
charges such as mass and angular momentum. Nevertheless,
because of the appearance of the G\"{o}del parameter, the naive application
of some traditional approaches breaks down in the computation of those quantities
\cite{KlVan}. To our knowledge, so far the first successful application towards
the neutral rotating G\"{o}del-type black holes has been made by the method proposed by
Barnich, Brandt and Comp\`{e}re \cite{BarnichB,Barnich,BarnichC,BCintegC}
in the context of the five-dimensional minimal supergravity \cite{BarCCC}
(We shall refer to this method as BBC approach for short. Up to now it has been widely used
to compute the conserved quantities associated with asymptotic symmetries). Afterwards
this approach was extended to the three-dimensional G\"{o}del-type
black holes of Einstein-Maxwell theory \cite{BaBCG3d}, as well as the
charged counterpart \cite{WuPtherG}.
However, for the squashed-horizon generalization
of the G\"{o}del-type black holes, the counterterm method is also feasible.
This is attributed to the fact that the squashing transformation essentially
modifies the asymptotic structure although it reserves the G\"{o}del
parameter \cite{StelSWsquashG}. As a consequence, despite the success of the
BBC approach, it is still desirable to clarify whether other methods may work
well for the G\"{o}del-type solutions for the purpose of providing multi-angle
analysis on their conserved quantities, as well as verifying the universality
of the methods on the definition of the conserved charges.

The Abbott-Deser-Tekin (ADT) formalism \cite{AbbottD,AbbottD2,DeserT,DeserT2,DeserT3},
constructed out of the conserved current obtained in virtue of the linearized
perturbation for the expression of gravitational field equation in a fixed reference
background of flat or (A)dS spacetime, has made some progress towards the
computation scheme for conserved quantities of asymptotically flat or (A)dS solutions in a variety of
theories of gravity. Since the relevant background metric is required to be a vacuum solution of the field
equation, both the current and potential in the ordinary ADT formalism are on-shell.
However, it was suggested that the on-shell constraint of the reference background is able to be relieved
\cite{KimKY,BouchClem,MouClGuL}. In particular, this was systematically elaborated
by Kim, Kulkarni and Yi in Ref. \cite{KimKY}. Here the authors constructed the current
and potential associated with the diffeomorphism symmetry of arbitrary background
spacetimes within the framework of generic covariant pure gravity theories,
and they further presented an off-shell prescription to
the ordinary ADT formalism. In practice, such a modification
leads to the advantage that it becomes more operable to derive the potential in terms of
the corresponding conserved
current. What is more, by contrast with the original ADT formalism that
involves no matter fields in the perturbation of fields,
it is of great convenience to incorporate the contributions from the matter fields
within the off-shell formulation \cite{CiteHJPY,JJPengpform} along the
lines of the well-known covariant
phase space approach, put forward by Wald and his collaborators \cite{LeeWald,IyerWald,IWaldentro}.
Apart from this, the manner of the perturbation for the fields in the off-shell generalization
has been modified as the same as that in the BBC approach. Consequently, in some sense,
the off-shell generalized ADT formalism could also be viewed as the cousin of both the
BBC and phase space approaches.
Till now, there have existed a series of developments and applications of the off-shell
ADT formulation in a range of gravity theories
regardless of the asymptotic behavior of the spacetime metrics
\cite{ABCHJ,Kulasa,KBhaMaj,KBADMaj,AdaSeST,HerVas,JJPsquaK,PXCsquaK,JPKeAd,SeAda,SetAdam,Pstrith,JJPengPLB,CiteHPY,Wuli}.

In the present work, we will attempt to find out whether the original ADT formalism together
with its off-shell generalization can produce the physical mass and angular momentum of the
charged rotating G\"{o}del-type black holes found in \cite{GodelBHWuGi}, just like the BBC
approach does \cite{WuPtherG}.
As we shall explicitly demonstrate below, within the context of supergravity theories with the
Einstein-Maxwell-Chern-Simons Lagangian in arbitrary odd spacetime dimensions, the off-shell
generalized ADT formalism including the contribution from the U(1) gauge field is
consistent with the BBC approach for the exact Killing vectors. Therefore, the off-shell
formulation can be naturally adopted as another suitable candidate to serve as
the definition of the conserved charges for the G\"{o}del-type
black holes. By contrast, the original ADT method fails as the case with respect to
the neutral G\"{o}del-type black holes \cite{KlVan} provided that the gauge field
is neglected. To tackle this issue, it seems to be a good strategy to follow closely the spirit of the
original ADT method to associate the contribution from the gauge field with the conserved
quantities. But we still have to face the important challenge of looking for appropriate reference background,
which is just one of the involved procedures for the conventional ADT formalism. In practice,
our consequences further demonstrate that the G\"{o}del-type black holes are good examples to
show how the matter fields to affect the conserved charges.

The remainder of this work is structured as follows. In section \ref{two}, we will give a
brief derivation of the off-shell potentials as well as the formula for the conserved charges
associated with the bosonic supergravity theories in arbitrary odd dimensions. In section \ref{three},
we will investigate the mass and angular momentum of the five-dimensional charged rotating
G\"{o}del-type black holes through the off-shell generalized ADT formalism. In section
\ref{four}, their mass and angular momentum will be reconsidered
by strictly following the original ADT formulation. The last section contains some conclusions.

\section{The formula of the conserved charges for $(2n+1)$-dimensional
supergravity theories}\label{two}

In this section, taking into consideration of the contribution from the matter field, we shall
derive the off-shell Noether and ADT potentials associated with $(2n+1)$-dimensional bosonic
supergravity theories along the lines of the work \cite{JJPengpform},
which recently developed the off-shell generalized ADT formalism in \cite{KimKY,CiteHJPY}.
Then a general formula of conserved charges for these theories will be
presented on basis of the potentials.

Without loss of generality, let us take as the start point the supergravity theory.
This is the extension of general relativity with an additional Abelian gauge field
and described by the Einstein-Maxwell-Chern-Simons Lagrangian with the cosmological
constant $\Lambda$ in odd spacetime dimensions $D=2n+1$. We write down the following
general expression:
\bea
\mathcal{L}_{EMCS}&=&\sqrt{-g}(L_{(gr)}+L_{(em)}+L_{(cs)}) \, , \nn \\
L_{(gr)}&=&R-2\Lambda \, , \quad L_{(em)}=\alpha F_{\mu\nu}F^{\mu\nu}
\, , \nn \\
L_{(cs)}&=&\beta \epsilon^{\gamma\mu_1\nu_1\cdot\cdot\cdot\mu_n\nu_n}
A_\gamma F_{\mu_1\nu_1}\cdot\cdot\cdot F_{\mu_n\nu_n}
\, , \label{LagofEinMCS}
\eea
where the parameters $\alpha$ and $\beta$ are arbitrary coupling constants, and the
completely antisymmetric Levi-Civita tensor
$\epsilon^{\mu_1\cdot\cdot\cdot\mu_{D}}$ in $D$ dimensions is defined through
$\sqrt{-g}\epsilon^{\mu_1\cdot\cdot\cdot\mu_{D}}=
-D!\delta_0^{[\mu_1}\cdot\cdot\cdot\delta_{D-1}^{\mu_D]}$.
Varying the Lagrangian (\ref{LagofEinMCS}) with respect to the metric $g_{\mu\nu}$
and the gauge field $A_\mu$, one obtains the equations of motion, given by
\bea
0&=&
R_{\mu\nu}-\frac{1}{2}g_{\mu\nu}R +g_{\mu\nu}\Lambda+2\alpha \Big(F_{\mu\rho}F_\nu^{~\rho}
-\frac{1}{4}g_{\mu\nu}F^2\Big) \, ,   \nn \\
0&=&4\alpha \nabla_\nu F^{\mu\nu}
+(n+1)\beta \epsilon^{\mu\mu_1\nu_1\cdot\cdot\cdot\mu_{n}\nu_{n}}
F_{\mu_1\nu_1}\cdot\cdot\cdot F_{\mu_{n}\nu_{n}} \, , \label{EOMofEinMCS}
\eea
as well as the surface term $\Theta^\mu$, which can be expressed as
the decomposition
\be
\Theta^\mu =\Theta_{(gr)}^\mu
+\Theta_{(em)}^\mu +\Theta_{(cs)}^\mu
\,  \label{SurfTofEinMCS}
\ee
in terms of the
quantities $\Theta_{(gr)}^\mu$, $\Theta_{(em)}^\mu$ and $\Theta_{(cs)}^\mu$,
which are read off as
\bea
\Theta_{(gr)}^\mu&=&
2\nabla^{[\sigma} h^{\mu]}_{\sigma} \, , \qquad
\Theta_{(em)}^\mu=4\alpha F^{\mu\nu} a_\nu \, , \nn \\
\Theta_{(cs)}^\mu&=&2n\beta
\epsilon^{\gamma\mu_1\nu_1\cdot\cdot\cdot\mu_{n-1}\nu_{n-1}\mu\nu}
A_\gamma F_{\mu_1\nu_1}\cdot\cdot\cdot F_{\mu_{n-1}\nu_{n-1}}a_\nu
\, , \label{SurfTofEinMCS2}
\eea
where $h_{\rho\sigma}=\delta g_{\rho\sigma}$ and $a_\nu=\delta A_\nu$ denote the perturbations
of the metric and $U$(1) gauge field, respectively. As usual, the spacetime indices of the
perturbations for the fields are lowered and raised using the background metric tensor $g_{\mu\nu}$
and its inverse $g^{\mu\nu}$ respectively.

Now, we focus our attention upon the derivations of the off-shell ADT potential associated with
the Lagrangian (\ref{LagofEinMCS}) on basis of its underlying diffeomorphism symmetry.
Specifically, supposing that the spacetime admits a series of symmetries
generated by the Killing vector $\xi^\mu$, we start with the off-shell Noether potential
$K^{\mu\nu}$ corresponding to this vector. In light of the consequences for the potentials of general
$p$-form gauge fields given in \cite{JJPengpform}, the potential $K^{\mu\nu}$ is expressed as the linear
combination of the three ones $K^{\mu\nu}_{(gr)}$, $K^{\mu\nu}_{(em)}$ and $K^{\mu\nu}_{(cs)}$,
that is,
\be
K^{\mu\nu}=K^{\mu\nu}_{(gr)}+K^{\mu\nu}_{(em)}+K^{\mu\nu}_{(cs)}
\, , \label{NoethPKEinMCS}
\ee
in which the potentials $K^{\mu\nu}_{(gr)}$,
$K^{\mu\nu}_{(em)}$ and $K^{\mu\nu}_{(cs)}$ correspond to the contributions from the
Einstein-Hilbert Lagrangian $\sqrt{-g}L_{(gr)}$, the electromagnetic field Lagrangian
$\sqrt{-g}L_{(em)}$ and the Chern-Simons term $\sqrt{-g}L_{(cs)}$ respectively,
defined by
\bea
K^{\mu\nu}_{(gr)}&=&2\nabla^{[\mu}\xi^{\nu]}\, , \qquad
K^{\mu\nu}_{(em)}=-4\alpha \vartheta F^{\mu\nu}  \, , \nn \\
K^{\mu\nu}_{(cs)}&=&-2n\beta\vartheta
\epsilon^{\gamma\mu_1\nu_1\cdot\cdot\cdot\mu_{n-1}\nu_{n-1}\mu\nu}
A_\gamma F_{\mu_1\nu_1}\cdot\cdot\cdot F_{\mu_{n-1}\nu_{n-1}}
\, , \label{Kgremcs}
\eea
where the scalar $\vartheta=A_\sigma \xi^\sigma$.
As usual, the off-shell Noether current corresponding to $K^{\mu\nu}$ is expressible as
$J^\mu=\nabla_\nu K^{\mu\nu}$. Furthermore, with the help of the surface term
$\Theta^\mu$ and the potential $K^{\mu\nu}$, it is feasible
to define the off-shell ADT potential $Q^{\mu\nu}$ involved in the Lagrangian (\ref{LagofEinMCS})
\cite{KimKY,CiteHJPY,JJPengpform,BouchClem,MouClGuL}. Like before, for convenience, such a potential
consists of  $Q^{\mu\nu}_{(gr)}$, $Q^{\mu\nu}_{(em)}$ and $Q^{\mu\nu}_{(cs)}$ three
ingredients, taking the form
\be
Q^{\mu\nu}=Q^{\mu\nu}_{(gr)}+Q^{\mu\nu}_{(em)}+Q^{\mu\nu}_{(cs)} \, .
 \label{ADTPEinMCS}
\ee
In the above equation, according to the works \cite{CiteHJPY,JJPengpform}, all the potentials
are defined through $Q^{\mu\nu}_{N}=\delta\big(\sqrt{-g}K_{N}^{\mu\nu}\big)/\big(2\sqrt{-g}\big)
-\xi^{[\mu}\Theta_{N}^{\nu]}$, where $N$'s refer to $(gr)$, $(em)$ and $(cs)$. Consequently,
the potentials $Q^{\mu\nu}_{(gr)}$ and $Q^{\mu\nu}_{(em)}$,
associated with the Lagrangians $\sqrt{-g}L_{(gr)}$ and $\sqrt{-g}L_{(em)}$
respectively, are presented by
\bea
Q^{\mu\nu}_{(gr)}
&=&\xi_\rho \nabla^{[\mu}h^{\nu]\rho}
-h^{\rho[\mu} \nabla_{\rho}\xi^{\nu]}
+\frac{1}{2}h \nabla^{[\mu}\xi^{\nu]}
-\xi^{[\mu} \nabla_\rho h^{\nu]\rho}
+\xi^{[\mu} \nabla^{\nu]}h \, , \nn \\
Q^{\mu\nu}_{(em)}
&=&-\alpha\Big[\vartheta\Big(2f^{\mu\nu}-h F^{\mu\nu}
+6h_{\sigma}^{[\sigma}F^{\mu\nu]}\Big)
+6a_\sigma\xi^{[\sigma}F^{\mu\nu]}\Big]
\, , \label{ADTPEinMax}
\eea
in which $h=g^{\mu\nu}h_{\mu\nu}$ and
$f_{\mu\nu}=2\partial_{[\mu}a_{\nu]}$, while the contribution from the Chern-Simons
term $\sqrt{-g}L_{(cs)}$ is
\bea
Q^{\mu\nu}_{(cs)}
&=&\tilde{Q}^{\mu\nu}_{(cs)}+\nabla_\rho U^{\rho\mu\nu}
\, . \label{ADTPotinCS}
\eea
Here the three-form $U^{\rho\mu\nu}$ and the two-form $\tilde{Q}^{\mu\nu}_{(cs)}$
are further given by
\bea
U^{\rho\mu\nu}&=&2n(n-1)\beta\vartheta
\epsilon^{\rho\mu\nu\mu_1\nu_1\cdot\cdot\cdot\mu_{n-1}\nu_{n-1}}
A_{\mu_1} a_{\nu_1} F_{\mu_2\nu_2}\cdot\cdot\cdot F_{\mu_{n-1}\nu_{n-1}} \, , \nn \\
\tilde{Q}^{\mu\nu}_{(cs)}&=&-n(n+1)\beta\vartheta
\epsilon^{\mu\nu\gamma\mu_1\nu_1\cdot\cdot\cdot\mu_{n-1}\nu_{n-1}}
a_{\gamma} F_{\mu_1\nu_1}\cdot\cdot\cdot F_{\mu_{n-1}\nu_{n-1}}
\, . \label{UandCSpote}
\eea
One can immediately observes that $U^{\rho\mu\nu}$ makes no contribution to the
off-shell Noether current
$\mathcal{J}^\mu=\nabla_\nu Q^{\mu\nu}$. In contrast to the potential defined
in terms of the original ADT formalism, here the potential $Q^{\mu\nu}$ incorporates
the contribution from the matter field and it does not demand that the background
fields have to be on-shell. Besides, the perturbations of all the fields in
$Q^{\mu\nu}$, determined by the variation of the solution parameters, take a different
manner from the ones in the usual ADT potential, which leads to that the off-shell
generalized ADT formalism can neglect the key step to seek appropriate reference backgrounds.
Apart from the off-shell generalized ADT formalism, it is worthwhile mentioning that the
phase space approach \cite{LeeWald,IyerWald,IWaldentro} can also yield the off-shell
Noether potential $K^{\mu\nu}$ in Eq. (\ref{NoethPKEinMCS}) as well
as the ADT potential $Q^{\mu\nu}$ in Eq. (\ref{ADTPEinMCS}), for instance,
the potential $K^{\mu\nu}$ coincides with the one obtained through the phase space method in \cite{RogEMCS}.
However, the currents defined through the two approaches differ from each other by the terms
proportional to the expressions for the equations of motion.

Next, we switch to the comparison between the off-shell generalized ADT potential $Q^{\mu\nu}$
and the superpotential $k^{\mu\nu}$ presented in \cite{BarCCC}, which was derived in terms
of the BBC approach. We find that both of
them differ from each other only by the divergence of $ U^{\rho\mu\nu}$, namely,
\be
Q^{\mu\nu}=k^{\mu\nu}+\nabla_\rho U^{\rho\mu\nu} \, .
\ee
In this regard, one can conclude that both the conserved potentials $Q^{\mu\nu}$
and $k^{\mu\nu}$ are equivalent when the spacetime symmetry is generated by the
exact Killing vector. Further combined with the fact that the manner of
the perturbations for the fields involved in the off-shell generalized ADT method
is as same as the one in the framework of the BBC approach, one observes that they
are equivalent for the supergravity theories described by the
Lagrangian (\ref{LagofEinMCS}). In other words, the off-shell generalized ADT method
could provide an alternative way to derive the superpotential in the BBC approach.

Finally, supposed that there exists a subregion given by a $(D-1)$-dimensional
hypersurface $\Sigma$ with the boundary $\partial\Sigma$, we are able to define
the conserved charge $\mathcal{Q}$ associated with this subregion in such a way that
\be
\delta \mathcal{Q}=\frac{1}{8\pi} \int_{\partial\Sigma} Q^{\mu\nu}
d\Sigma_{\mu\nu}
\,  \label{dQdefineAn}
\ee
in terms of the potential $Q^{\mu\nu}$ within the context of the $D$-dimensional
supergravity theories described by the Lagrangian (\ref{LagofEinMCS}). In the above
equation, $d\Sigma_{\mu\nu}=\frac{1}{2}\frac{1}{(D-2)!}
\epsilon_{\mu\nu\mu_1\mu_2\cdot\cdot\cdot\mu_{(D-2)}}dx^{\mu_1}\wedge\cdot\cdot\cdot
\wedge dx^{\mu_{(D-2)}}$. One can thus obtain the conserved
quantity $\mathcal{Q}$ by integrating Eq. (\ref{dQdefineAn}) out.

\section{Mass and angular momentum in the off-shell generalized ADT formulation}\label{three}
In this section, we shall pay our attention to carrying out the off-shell generalized ADT formalism
to investigate the mass and angular momentum of the rotating charged black hole
in the G\"{o}del-type universe found in the work \cite{GodelBHWuGi}, which is an exact solution of the
equations of motion given by Eq. (\ref{EOMofEinMCS}) within the framework of the
five-dimensional Einstein-Maxwell-Chern-Simons supergravity theory,
described by the Lagrangian (\ref{LagofEinMCS}) with $\Lambda=0$, $\alpha=-1$ and
$\beta=-2\sqrt{3}/9$. The line element for these 4-parameter black holes takes the
general form
\bea
ds^2&=&-\frac{\big(Udt+H\sigma_3\big)^2}{U}
+\frac{dr^2}{V}
+\frac{r^2}{4}\big(d\theta^2+\sin^2\theta d\psi^2\big)
+\frac{r^2V\sigma_3^2}{4U} \, , \nn \\
A&=&\frac{\sqrt{3}q}{2r^2}dt
+\frac{\sqrt{3}}{2}W\sigma_3
\,  \label{5DGodelBH}
\eea
with the quantities $U$ and $W$ given by
\be
U=1-\frac{2m}{r^2}+\frac{q^2}{r^4} \, , \quad
W=jr^2+2jq-\frac{aq}{2r^2}
\, , \label{UVHdefin}
\ee
while $H$ and $V$ can be expressed through $U$ and $W$ as following:
\bea
H&=&W+jq-\frac{a}{2}(U-1)
\, ,  \nn \\
V&=&U+\frac{8j(m+q)(a+2jm+4jq)}{r^{2}} \nn \\
&&-\frac{2(a+2jq)[qa-ma+2jq(m+3q)]}{r^{4}}
\, . \label{UVHdefin2}
\eea
In Eq. (\ref{5DGodelBH}), the left invariant form $\sigma_3=d\phi+\cos\theta d\psi$.
The parameter $j$ characterizes the scale of the G\"{o}del background and plays the
role for the rotation of the universe, while we shall see in a moment that all the
integral constants $m$, $a$ and $q$ are related to the mass, angular momentum and
electric charge, respectively. Besides, the Euler angles $\theta$, $\phi$ and $\psi$
range from $0<\theta<\pi$, $0<\phi<4\pi$ and $0<\psi<2\pi$ respectively.
The solution (\ref{5DGodelBH}) behaves asymptotically as the five-dimensional rotating
G\"{o}del-type universe, which provides an ideal platform to test the various traditional
properties of gravity theories 
\cite{PouKoSa,EiReDy,LisupeGodl,KoZhisrG,SeKaHc,PWhidSym,CWJSem,PeWuKCF}.
As a matter of fact,
the metric (\ref{5DGodelBH}) can cover several well-known solutions in the
literature on the choices of appropriate parameters. For instance, when $m,a,q=0$,
it becomes the maximally supersymmetric analogues of the five-dimensional
G\"{o}del universe \cite{GGHPRGod}. When the parameter
$q$ or $j$ vanishes, it returns to the one of the neutral rotating G\"{o}del
black hole found in \cite{GodelGimH} or the charged black holes with two equal rotations
in \cite{CvetiY,ChongCLP}. What is more, in the absence of both
the parameters $j$ and $q$, the solution (\ref{5DGodelBH}) represents that of the rotating
black holes satisfying the five-dimensional vacuum Einstein equation.

Now we concentrate on dealing with the mass $M$ of the rotating charged G\"{o}del-type black hole
(\ref{5DGodelBH}). To do this, it is necessary to identify the perturbations of all the fields first.
According to the off-shell generalized ADT formulation, the fluctuations $h_{\mu\nu}$ of
the spacetime metric and the ones $a_\mu$ of gauge field could be determined by the
infinitesimal changes of all the three solution parameters $(m,a,q)$, that is,
\be
m\rightarrow m+\delta m \, , \quad
a\rightarrow a+\delta a \, , \quad
q\rightarrow q+\delta q
\, . \label{Flucofmaq}
\ee
Here it is worth noting that the G\"{o}del parameter $j$ can not be adopted to perturbate the
gravitational and gauge fields, otherwise the mass and angular momentum are non-integrable.
To gain the mass $M$, we have to evaluate the $(t,r)$ component of the off-shell ADT potential
$\sqrt{-g}Q^{\mu\nu}\big(\xi_{(t)}\big)$ associated with the timelike Killing vector
$\xi^\mu_{(t)}=(-1,0,0,0,0)$, which is read off as
\bea
\sqrt{-g}Q^{tr}\big(\xi_{(t)}\big)&=&
\sqrt{-g}\big(Q^{tr}_{(gr)}+Q^{tr}_{(em)}+Q^{tr}_{(cs)}\big)
\, , \label{OffshQtr}
\eea
where
\bea
\sqrt{-g}Q^{tr}_{(gr)}&=&3j^2r^2\sin\theta \delta q
-\frac{\sin\theta}{8}\big[(4aj+72qj^2+64mj^2-3)\delta m
+2j(2m+5q)\delta a\nn \\
&&+2j(28jq+60mj+5a)\delta q\big]+\mathcal{O}\big(r^{-2}\big) \, , \nn \\
\sqrt{-g}Q^{tr}_{(em)}&=&-3j^2r^2\sin\theta \delta q
+\frac{3}{4}j\sin\theta\big[q\delta a+(a+8jm-12jq)\delta q\big]
+\mathcal{O}\big(r^{-2}\big) \, , \nn \\
\sqrt{-g}Q^{tr}_{(cs)}&=&
\frac{3qW\sin\theta}{r^2}\delta W
\, . \label{ComoffQtr}
\eea
According to Eq. (\ref{ComoffQtr}), we find that the contribution from the Lagrangian
for Maxwell's equations can not be neglected, otherwise the charge is divergent
when $r\rightarrow \infty$. Moreover, the Chern-Simons term contributes to the
total charge as well. This is different from the cases in \cite{JJPsquaK,Wuli}.
Therefore, a naive application of the off-shell generalized
ADT foramlism without the contributions from the matter fields may fail to yield
physical mass of the rotating charged G\"{o}del black holes. Further making use of
the integration of the formula (\ref{dQdefineAn}) for the conserved charges,
expressed as
\be
\mathcal{Q}=\frac{1}{8\pi} \int_{\partial\Sigma}
\int^{g}_{\bar{g}}\int^{A}_{\bar{A}}
Q^{\mu\nu}d\Sigma_{\mu\nu}
\, , \label{IntegQ}
\ee
in which we opt for the background metric $\bar{g}=g|_{m,a,q=0}$ and the reference
gauge field $\bar{A}=A|_{m,a,q=0}$, we arrive at
\be
M=\pi\Big[\frac{3}{4}m-2(m+q)(4m+5q)j^2-aj(m+q)\Big]
\, . \label{MassofGoBH}
\ee
Here the mass $M$ exactly agrees with the one through the BBC approach and it has been
shown that $M$ satisfies the first law of black hole thermodynamics in \cite{GodelBHWuGi,WuPtherG}.
In the absence of the electric parameter $q$, $M$ returns to the result given in
\cite{BarCCC}.

Next, we move on to evaluate the angular momentum $J_\phi$ along the $\phi$ direction
in the off-shell generalized ADT formalism. As a consequence of the vanishing for the
$(t,r)$ component of the term $\xi_{(\phi)}^{[\mu}\Theta^{\nu]}$ in the potential
$Q^{\mu\nu}\big(\xi_{(\phi)}\big)$ given by Eq. (\ref{ADTPEinMCS}), where
$\xi^\mu_{(\phi)}=(0,0,0,1,0)$ is the relevant spacelike Killing vector, it is convenient
for us to compute $J_\phi$ in terms of the following generalized formulation for
the standard definition of the Komar angular momentum:
\be
J=\frac{1}{16\pi} \Big[\int_{\partial\Sigma} K^{\mu\nu}
d\Sigma_{\mu\nu} \Big]^{(g,A)}_{(\bar{g},\bar{A})}
\, , \label{JinKomar}
\ee
In comparison with the general formula (\ref{dQdefineAn}),
the merit of the formula (\ref{JinKomar}) is that it eliminates the step of perturbating
both the gravitational and gauge fields. Hence the calculations are significantly
simplified. It should be emphasized that the Noether potential
$K^{\mu\nu}$ here is different from the superpotential in the ordinary Komar integral by
incorporating the contribution from the matter field. As an application to the rotating
charged G\"{o}del black hole (\ref{5DGodelBH}), we calculate the $(t,r)$ component of
the Noether potential $K^{\mu\nu}$ associated with the Killing vector $\xi^\mu_{(\phi)}$,
consisting of the contribution from the gravitational field $K^{tr}_{(gr)}$, given by
\bea
\sqrt{-g}K^{tr}_{(gr)}&=&-j^3\big(r^2+6q\big)r^4\sin\theta
-\frac{3}{4}j\sin\theta\big[8q(m+3q)j^2+2aj(2m+q)-q\big]r^2 \nn \\
&&-\frac{a}{4}\sin\theta\big[8j^2\big(2m^2+5mq-4q^2\big)+4aj(m-q)+q-2m\big]
+\mathcal{O}\big(r^{-2}\big)
\, , \label{EMCSKgr}
\eea
in combination with the ones from the $U$(1) gauge field $K^{tr}_{(em)}$ and $K^{tr}_{(cs)}$,
expressed as
\bea
\sqrt{-g}K^{tr}_{(em)}&=&3j^3\big(r^2+6q\big)r^4\sin\theta
+\frac{3}{4}j\sin\theta\big[8q(m+7q)j^2+2aj(2m-q)-q\big]r^2 \nn \\
&&+\frac{3}{2}jq\sin\theta\big[8q(m+3q)j^2+4aj(m-2q)-q\big]
+\mathcal{O}\big(r^{-2}\big) \, , \nn \\
\sqrt{-g}K^{tr}_{(cs)}&=&-2W^3\sin\theta
\, . \label{EMCSKem}
\eea
According to Eqs. (\ref{EMCSKgr}) and (\ref{EMCSKem}), because of the existence of the
G\"{o}del parameter $j$, if one merely let the Noether potential
$K^{\mu\nu}_{(gr)}$ with respect to the gravitational field enter the definition for
the conserved charge of the charged rotating G\"{o}del black holes as the ordinary Komar integral
does, one will
observe that the angular momentum is divergent at infinity. Consequently, to get
round the problem of divergence it is of considerable necessity to take
into account the contribution from the Maxwell Lagrangian $\sqrt{-g}L_{(em)}$ as
well as that from the Chern-Simons term $\sqrt{-g}L_{(cs)}$, even for the neutral
rotating G\"{o}del black holes \cite{GodelGimH}.
Further substituting the $(t,r)$ component of the total off-shell Noether potential
$\sqrt{-g}K^{tr}\big(\xi_{(\phi)}\big)=\sqrt{-g}\big(K^{tr}_{(gr)}+K^{tr}_{(em)}+K^{tr}_{(cs)}\big)$
into the formula (\ref{JinKomar}), we gain the angular momentum $J_\phi$ along
the $\phi$ direction
\bea
J_\phi&=&\frac{\pi}{4}\big[a(2m-q)-4aj(m-q)(a+4jm+8jq) \nn \\
&&-6jq^2+16(3m+5q)q^2j^3\big]
\, , \label{AnguJphi}
\eea
which coincides with the one through the BBC approach and is the desired
thermodynamical quantity in fulfillment of the requirements for the first
law of black hole thermodynamics \cite{GodelBHWuGi,WuPtherG}. Here the
angular momentum $J_\phi$ covers the one in \cite{BarCCC}
as a special situation.

\section{Mass and angular momentum via the original ADT formalism}\label{four}

In this section, we shall comply fully with the rules of the original ADT formalism
\cite{AbbottD,AbbottD2,DeserT,DeserT2,DeserT3}
to modify the potential by including the contribution from the gauge field in the context
of the $(2n+1)$-dimensional Einstein-Maxwell-Chern-Simons supergravity theories.
In comparison with the off-shell generalized ADT formulation, then we will take into
account the mass and angular momentum of the five-dimensional charged rotating G\"{o}del
black holes in terms of the generalized potential, accompany with a detailed analysis
of different reference backgrounds.

Let us start with the potential within the framework of the original ADT formalism. According
to this formulation, we observe that the usual potential contains no contributions from the matter
fields, such as the gauge field and the scalar field, since it is assumed that all the matter
fields decrease quite fast so that their contributions to the total conserved charges can
be ignored. However, as is demonstrated in the previous section, this does not hold true at leat for the
G\"{o}del-type black holes, let alone the black holes in Horndeski theory \cite{JJPengPLB,SBiHorden}.
Therefore, for the completeness and universality of this formalism, it is of great necessity to
simultaneously take into consideration of the effects of the matter fields. Like in \cite{JJPsquaK},
here we incorporate the contribution from the gauge field $A_\mu$ within the potential
associated with the Lagrangian (\ref{LagofEinMCS}) in full analogy with the rules of the original
ADT formulation. Such an obviously extended potential $\bar{Q}^{\mu\nu}$ can be chosen as the one
in Eq. (\ref{ADTPEinMCS}) with the modification that both the background metric and gauge
field $(\bar{g}_{\mu\nu},\bar{A}_{\mu})$ are required to be on-shell and fixed, as well
as the perturbations of the gravitational and gauge fields $(h_{\mu\nu},a_\mu)$ are replaced
by the differences between the given fields and the fixed reference ones, that is,
\be
h_{\mu\nu}=g_{\mu\nu}-\bar{g}_{\mu\nu} \, , \qquad
a_\mu=A_\mu-\bar{A}_\mu
\, , \label{vargA}
\ee
instead of the fluctuations of the solution parameters. In accordance with
the notations in the usual ADT formulation, $\bar{Q}^{\mu\nu}$ is expressed as
\be
\bar{Q}^{\mu\nu}=
Q^{\mu\nu}(g\rightarrow\bar{g},A\rightarrow\bar{A};\nabla\rightarrow\bar{\nabla})
\, . \label{ModifiedQ}
\ee
Here $\bar{\nabla}$ denotes the covariant derivative operator of the connection
with respect to the fixed background metric tensor $\bar{g}_{\mu\nu}$. Our generalized
potential (\ref{ModifiedQ}) containing the
contribution from the gauge field is completely constructed in spirit
of the original ADT formalism. Hence the conserved charges defined in terms of
this potential could be regarded as a simple and apparent generalization to the ordinary ADT
formula in the minimal supergravity theories.

With the modified ADT potential (\ref{ModifiedQ}) in hand, we proceed to calculate the mass
and angular momentum of the charged rotating G\"{o}del black holes. As is known,
the ADT formalism
is background-dependent. Thus an appropriate reference background, which may be
non-unique, plays an important role in the calculations. In the literature, a common
way of obtaining that is to let all or selected solution
parameters vanish. By following this, a natural reference background
$(\bar{g}_{(1)},\bar{A}_{(1)})$ for the charged rotating G\"{o}del
black holes could be set as the maximally supersymmetric
G\"{o}del universe \cite{GGHPRGod}, with the form
\bea
d\bar{s}^2_{(1)}&=&ds^2\big|_{m,a,q=0}
\nn \\
&=&-\big(dt+jr^2\sigma_3\big)^2+dr^2
+\frac{r^2}{4}\big(d\theta^2+\sin^2\theta d\psi^2+\sigma_3^2\big)
\, , \nn \\
\bar{A}_{(1)}&=&c_tdt+c_\phi d\phi+c_\psi d\psi +\frac{\sqrt{3}}{2}jr^2\sigma_3^2
\, . \label{RefBack}
\eea
Here we preserve the arbitrary constants $c_t$, $c_\phi$ and $c_\psi$ for generality.
On the other hand, the manipulation to let all the three parameters disappear leads to a simpler background
gauge field. However, as we shall see below, their vanishing can not ensure the
convergence of the $(t,r)$ component of the potential $\sqrt{-\bar{g}}\bar{Q}^{\mu\nu}$
at infinity. One can verify that the background fields
$\bar{g}^{(1)}_{\mu\nu}$ and $\bar{A}^{(1)}_\mu$ in Eq. (\ref{RefBack}) are on-shell, that is,
they strictly satisfy the five-dimensional field equations with $\Lambda=0$, $\alpha=-1$ and
$\beta=-2\sqrt{3}/9$ given by Eq. (\ref{EOMofEinMCS}). Furthermore,
on the fixed reference metric $\bar{g}^{(1)}_{\mu\nu}$ and gauge field
$\bar{A}^{(1)}_\mu$ to compute $\sqrt{-\bar{g}}\bar{Q}^{tr}$, we obtain
\bea
\sqrt{-\bar{g}}\bar{Q}^{tr}\big(\xi_{(t)}\big)&=&j\sin\theta
\big[\sqrt{3}c_\phi-8c_t c_\phi/3-2\sqrt{3}c_t(2m+3q)^2j^3
+(2m+3q)^2j^3\big]r^2 \nn \\
&&+C\sin\theta+\mathcal{O}\big(r^{-2}\big)
\, , \label{Qbartr}
\eea
with the constant $C$ dependent upon the six parameters $c_t$, $c_\phi$, $m$,
$a$, $q$ and $j$, as well as
\bea
\sqrt{-\bar{g}}\bar{Q}^{tr}\big(\xi_{(\phi)}\big)&=&j^2\sin\theta
\big[c_\phi/\sqrt{3}+2(2m+3q)^2j^3\big]r^4+2j\sin\theta\times
\nn \\
&&\big\{4c_\phi^2/3+\sqrt{3}j\big[(2m+3q)^2j^2-2q/3\big] c_\phi
-m^2j^2\nn\\
&&+a(8m^2+8mq-3q^2)j^3-6(m+q)q^2j^4
\nn \\
&&-64(a+mj+2qj)(m+2q)(m+q)^2j^5
\nn \\
&&-4\big[4(am+aq)^2-m(2m+3q)^2\big]j^4\big\} r^2
+\mathcal{O}(1)
\, . \label{QbartrJ}
\eea
To guarantee that both the quantities
$\sqrt{-\bar{g}}\bar{Q}^{tr}\big(\xi_{(t)}\big)$ and
$\sqrt{-\bar{g}}\bar{Q}^{tr}\big(\xi_{(\phi)}\big)$ are convergent when the
radial coordinate $r$ goes to infinity, it is desired that the $r^2$ term in
the former, as well as the $r^4$ term in the latter, has to vanish, yielding
\be
c_t= \frac{\sqrt{3}}{2}\, , \quad
c_\phi= -2\sqrt{3}(2m+3q)^2j^3
\, . \label{ctphivalu}
\ee
However, it turns out that the values of the parameters $c_t$ and $c_\phi$ given by Eq. (\ref{ctphivalu})
are unable to make the $r^2$ term in $\sqrt{-\bar{g}}\bar{Q}^{tr}\big(\xi_{(\phi)}\big)$
disappear. This shows that the angular momentum along the $\phi$ direction still
suffers from the divergence. On the other hand, in such a case, although the convergent
$\sqrt{-\bar{g}}\bar{Q}^{tr}\big(\xi_{(t)}\big)$ gives rise to a finite value
for the mass, it differs from the one evaluated via the off-shell ADT formulation.

As is indicated in the above, the natural on-shell background metric and gauge field
$\big(\bar{g}^{(1)}_{\mu\nu},\bar{A}^{(1)}_\mu\big)$ in Eq. (\ref{RefBack}) fail to
produce both the meaningful mass and angular momentum of the charged rotating G\"{o}del
black holes because the divergence is unavoidable. To tackle such a problem, we have tried to look for
appropriate ones but find that the attempt is rather difficult
to achieve. However, if we just expect the reference background (\ref{RefBack})
to merely yield physical mass regardless of the angular momentum from the
pure mathematics perspective, as a matter of fact, this can be realised by adopting
appropriate $c_t$ and $c_\phi$ to render
the $r^2$ term in $\sqrt{-\bar{g}}\bar{Q}^{tr}\big(\xi_{(t)}\big)$ vanish as well as
$C$ take the value $M/(2\pi)$.

What is more, without the restriction to get the
mass and angular momentum on the same on-shell reference background, we are able to
put forward a more general reference background $(\bar{g}_{(2)},\bar{A}_{(2)})$ to compute
the mass in terms of the original ADT formulation from a mathematical point of view,
which is supposed to have the following form
\bea
d\bar{s}^2_{(2)}&=&ds^2\big|_{(m,a,q)\rightarrow(\bar{m},\bar{a},\bar{q})}
\, , \nn \\
\bar{A}_{(2)}&=&A|_{(m,a,q)\rightarrow(\bar{m},\bar{a},\bar{q})}
+\bar{c}_tdt+\bar{c}_\phi d\phi +\bar{c}_\psi d\psi
\, . \label{RefBack2}
\eea
Here the computations reveal that the parameter $\bar{c}_\psi$ actually makes no
contribution to the charges, so it can be set as $\bar{c}_\psi=0$ in advance.
However, the other five parameters $\bar{m}$, $\bar{a}$, $\bar{q}$,
$\bar{c}_t$ and $\bar{c}_\phi$ are desired to satisfy the following three conditions: \\
\textbf{\textrm{I}}. They guarantee that the reference line element $d\bar{s}^2_{(2)}$ has the same asymptotic
geometry as the original one $ds^2$ and the $(t,r)$ component of the ADT potential
$\sqrt{-\bar{g}}\bar{Q}^{tr}\big(\xi_{(t)}\big)$ is convergent when $r\rightarrow \infty$
so that the formula (\ref{dQdefineAn}) yields a finite quantity $\mathcal{M}$; \\
\textbf{\textrm{II}}. $\mathcal{M}=M-\bar{M}$, where the meaningful mass $M$
that fulfills the first law of thermodynamics is given by Eq. (\ref{MassofGoBH})
in the previous section, while $\bar{M}=M(m\rightarrow\bar{m},a\rightarrow\bar{a},q\rightarrow\bar{q})$
serves as the mass of the background spacetime $\bar{g}^{(2)}_{\mu\nu}$; \\
\textbf{\textrm{III}}. The mass $\bar{M}$ vanishes, namely, $\bar{M}=0$.\\
Here the three conditions to identify the reference background are very general. In fact, they
have been utilized in a range of applications for the original ADT formalism.

Actually, for the rotating charged G\"{o}del black hole (\ref{5DGodelBH}), apart from the special
background metric and gauge field $\big(\bar{g}^{(1)}_{\mu\nu},\bar{A}^{(1)}_\mu\big)$,
we indeed gain some other reference backgrounds appropriate for the computations of
the ADT mass by solving a set of equations for the five parameters in fulfillment
of the above-mentioned three conditions. Nevertheless, the results are rather complex.
In order to illustrate this, instead, we take into consideration of the situation for
the neutral rotating G\"{o}del-type black hole for simplicity. We find that the simplest
reference background is the one with
\bea
\bar{c}_t&=&\frac{\sqrt{3}}{6}\, , \quad \bar{c}_\phi=0\, , \quad
\bar{a}=\frac{3-32\bar{m}j^2}{4j}
\, ,\nn \\
\mathcal{M}&=&-\frac{\pi}{4}\big[64j^4\bar{m}^3-8(16mj^2-3)j^2\bar{m}^2
+(8mj^2+3)(8mj^2-1)\bar{m} \nn \\
&&+m(24mj^2+8ja-3)\big]
\, . \label{barmaofNeurGod}
\eea
Solving the equation $\mathcal{M}=M(q=0)$, where $M(q=0)=\pi m(3-32mj^2-4ja)/4$
is the mass of the neutral rotating G\"{o}del black hole, one can identify the parameter
$\bar{m}$. It should be emphasized that the parameters $(\bar{m}, \bar{a}, \bar{q})$ in
the background metric $\bar{g}^{(2)}_{\mu\nu}$ have no solutions fulfilling the three
conditions (\textbf{\textrm{I}}-\textbf{\textrm{III}}) provided that the generalized potential
(\ref{ModifiedQ}) lacks the contribution from the gauge field $A_\mu$ or both the
parameters $\bar{c}_t$ and $\bar{c}_\phi$ are naively set as $\bar{c}_t,\bar{c}_\phi=0$.
This implies that the mass is obtained at the expense of modifying the asymptotic
structure of the gauge field $A_\mu$. Besides, one can immediately conclude that the
original ADT method without consideration of the matter fields is insufficient for
producing the mass of the G\"{o}del-type black hole in a mathematical sense.

A remark is in order here. As has been demonstrated in the previous section, when
the off-shell generalized ADT potential $Q^{\mu\nu}$ in Eq. (\ref{ADTPEinMCS}) is adopted
to calculate the mass and angular momentum of the rotating charged G\"{o}del black holes,
one is able to obtain meaningful results. However, letting the modified potential
$\bar{Q}^{\mu\nu}$ in Eq. (\ref{ModifiedQ}) enter the formula for the ADT charges, we face the
difficulty of finding an appropriate reference background on which both the mass and
angular momentum can be simultaneously evaluated. This arises from the manner for
the fluctuations of the gravitational and gauge fields. Specifically, for the potential
$\bar{Q}^{\mu\nu}$, it is of significant importance to find in advance a proper fixed
reference background so that the perturbations of all the involved fields
can guarantee that the conserved charges are convergent at infinity. Nevertheless,
for the potential $Q^{\mu\nu}$, there is
no need to set the reference background in advance. All the fields fluctuate around
the background varying with the solution parameters and the integral of these parameters
gives rise to the charges. In this sense, the off-shell generalized ADT
formulation is more effective than the original one. However, from technical viewpoint,
the latter has the advantage for the avoidance of the integral to the conserved charge
$\delta \mathcal{Q}$ in the formula (\ref{dQdefineAn}).

\section{Conclusions}\label{five}

In this work, in order to present a concrete example to explicitly compare the
original ADT formalism \cite{AbbottD,AbbottD2,DeserT,DeserT2,DeserT3} and its
off-shell generalization \cite{KimKY}, as well as to stress the necessity to 
consider the contributions from the matter fields, according to both the 
formulations, we investigated the mass and angular momentum of the rotating 
charged black holes in the G\"{o}del-type universe \cite{GodelBHWuGi} within 
the framework of the five-dimensional Einstein-Maxwell-Chern-Simons 
supergravity theory.

As a starting point, we presented the off-shell generalized ADT potential 
$Q^{\mu\nu}$ (\ref{ADTPEinMCS}) in the context of $(2n+1)$-dimensional 
supergravity theories described by the Lagrangian (\ref{LagofEinMCS}). 
It was illustrated that the off-shell generalized ADT method is equivalent 
with the BBC approach. Next, the off-shell ADT formulation was applied to evaluate
the conserved charges of the rotating charged G\"{o}del black hole (\ref{5DGodelBH}).
We obtained the mass $M$ and angular momentum $J_\phi$ along the $\phi$ direction
given by Eqs. (\ref{MassofGoBH}) and (\ref{AnguJphi}) respectively. Finally, in comparison
with the off-shell generalized ADT formulation, we strictly followed the rules of the
original ADT formalism to incorporate the contribution from the $U$(1) gauge field within
the potential, yielding the generalized one $\bar{Q}^{\mu\nu}$ in Eq. (\ref{ModifiedQ}).
In terms of this potential, the original ADT formalism was utilized to the rotating
charged G\"{o}del black hole on the reference background $(\bar{g}_{(1)},\bar{A}_{(1)})$.
Unfortunately, we failed to obtain the mass and angular momentum on the same background.
Furthermore, regardless of the angular momentum, a general reference background
$(\bar{g}_{(2)},\bar{A}_{(2)})$ satisfying the three conditions
(\textbf{\textrm{I}}-\textbf{\textrm{III}}) was presented in Eq. (\ref{RefBack2}),
on which the original ADT formalism can produce the meaningful mass at the mathematical
level.

As what has been demonstrated, because of the appearance of the G\"{o}del 
parameter $j$, the gauge field's contribution involved in the potential 
$Q^{\mu\nu}$ or $\bar{Q}^{\mu\nu}$ has to be taken into consideration in 
order to get meaningful consequences. Therefore, apart from the black holes 
of Horndeski theory shown in \cite{JJPengPLB}, the G\"{o}del-type black 
holes can be chosen as ideal objects to display how the matter fields
to affect their mass and angular momentum. They are of great use for 
checking the universality of the approaches for the conserved charges.

The formalism with the modified potential $\bar{Q}^{\mu\nu}$,
which has incorporated the contribution from the gauge field in spirit 
of the original ADT formalism, might be extended to other black holes 
in the $(2n+1)$-dimensional Einstein-Maxwell-Chern-Simons supergravity 
theories, although this formulation is only a partial success for the 
five-dimensional (charged) rotating G\"{o}del-type black holes since the 
existence of the G\"{o}del parameter makes it rather difficult to seek 
for appropriate backgrounds. For instance, dealing with the mass and 
angular momenta of the general rotating charged black holes in 
five-dimensional gauged supergravity \cite{ChongCLP} in terms of the 
modified potential $\bar{Q}^{\mu\nu}$, we observe that the ADT formalism
yields the same meaningful results as the ones in \cite{MassCLP}, where 
the usual potential was adopted. This is attributed to the fast falloff 
of the gauge field as the radial coordinate goes to infinity. However, 
to see another example associated with the non-zero contributions from 
the gauge fields in the generalized potential, a sensible application 
might be that to the most general non-extremal rotating charged black 
holes in five-dimensional $U$(1)$^3$ gauged supergravity \cite{SQW3CBH}.

\section*{Acknowledgments}

This work was supported by the Natural Science Foundation of China under Grant
Nos. 11865006 and 11505036. It was also partially supported by the Technology
Department of Guizhou province Fund under Grant Nos. [2018]5769 and [2017]7349.

\end{document}